\documentclass[aps,prl,twocolumn,groupedaddress,showpacs]{revtex4} 
\usepackage{graphicx} 
  
\begin{document}  
  
\title{Post-Cotunnite Phase of TeO$_2$}  
  
\author{Gareth I. G. Griffiths}  
  
\affiliation {Theory of Condensed Matter Group, Cavendish Laboratory, 
J J Thomson Avenue, Cambridge CB3 0HE, United Kingdom} 
  
\author{R. J. Needs}  
  
\affiliation {Theory of Condensed Matter Group, Cavendish Laboratory, 
J J Thomson Avenue, Cambridge CB3 0HE, United Kingdom} 
 
\author{Chris J. Pickard} 
 
\affiliation {Department of Physics and Astronomy, University College 
London, Gower St, London WC1E 6BT, United Kingdom} 
 
\date{\today}  
  
\begin{abstract}  
We have used first-principles density-functional-theory methods with a
random-structure-searching technique to determine the structure of the
previously unidentified post-cotunnite phase of TeO$_2$.  Our
calculations indicate a transition from the cotunnite to
post-cotunnite phase at 130 GPa.  The predicted post-cotunnite
structure has $P2_1/m$ space group symmetry and its calculated x-ray
diffraction pattern is in good agreement with the available
experimental data.  We find that the cotunnite phase re-enters at
about 260 GPa.
 
\end{abstract}  
 
\pacs{62.50.-p,71.15.Nc,61.50.-f,91.60.Hg} 
 
 
\maketitle  
  
\section{Introduction}  
 
The majority of matter in the solar system is subject to pressures
above 10 GPa,\cite{ManjonE09} which motivates studies of materials
such as oxides at high pressures.  In the case of $AX_2$ compounds,
where $A$ and $X$ are, respectively, a divalent cation and halogen
atom or a tetravalent cation and oxygen atom, the general effect of
increasing the pressure is to distort the anion polyhedra and
eventually to increase the coordination number (CN).\cite{dewhurst01}
The highest CN observed in metal dioxides is in the PbCl$_2$-type
cotunnite structure with CN=9.  Metal dioxides with large cation radii
often form cotunnite phases at high pressures, such as TiO$_2$,
ZrO$_2$, HfO$_2$, CeO$_2$, PbO$_2$, PuO$_2$, UO$_2$, TbO$_2$,
TeO$_2$,\cite{kourouklis91} and ThO$_2$.\cite{lowther05}
The very important oxide SiO$_2$ is predicted to adopt the cotunnite 
structure above 690 GPa, which may be relevant to the study of 
extrasolar planets.\cite{UmemotoWA06} The hardest known oxide is the 
cotunnite structure of TiO$_2$, which has been synthesised at high 
pressures and recovered to ambient conditions.\cite{hardest} 
 
Materials that adopt the cotunnite structure are expected to transform
under additional applied pressure into post-cotunnite structures with
an accompanying increase in CN to 10 or more, as reported for some
dihalides.\cite{leger_halides} In reviewing the high pressure phases
of dioxides our attention was drawn to TeO$_2$ which, to the best of
our knowledge, is the only dioxide for which a transition to a
post-cotunnite phase has been observed.\cite{Sato} Sato \emph{et
al.}\cite{Sato} studied TeO$_2$ up to pressures of 150 GPa in a
diamond anvil cell.  X-ray diffraction data showed strong evidence for
a structural phase transition around 80-100 GPa, but the quality of
the data was insufficient to allow a determination of the structure of
the new phase, although the known post-cotunnite structures of
dihalides were eliminated.\cite{Sato} Identifying the post-cotunnite
structure of TeO$_2$ would further our understanding of dioxides at
high pressures.

TeO$_2$ is also an interesting material from the point of view of
fundamental science and technology.\cite{ChamparnaudMesjard20001499}
It has shown promise as a material for nonlinear optical devices,
usually in a glassy form but potentially from nanosize
crystals.\cite{LasbrugnasTMCBC05,costeteo2}

\section{Random structure searching}  
 
First-principles or \textit{Ab Initio} (AI) Density-Functional-Theory
(DFT) methods have been widely applied to materials at high pressures,
and have provided both confirmation of experimental results and
predictions of new phases and their properties.  DFT calculations have
given very accurate descriptions of the high-pressure phases of $sp$
bonded materials.\cite{MujicaRMN03} We have studied high-pressure
phases of TeO$_2$ using AI DFT methods combined with ``Random
Structure Searching'' (the AIRSS approach).\cite{silane_rss} This
approach has been used to predict high pressure phases which have
subsequently been found
experimentally,\cite{silane_rss,PickardN07_AlH3} and to predict new
high-pressure phases of materials such as hydrogen\cite{PickardN07_H2}
and ammonia.\cite{PickardN08_NH3}
  
The recipe for a random search commences by generating a set of
initial structures, for each of which a random unit cell is created
and renormalised to a reasonable volume, and the desired number of
each atomic species is randomly distributed throughout. Each of these
initial configurations is relaxed to a minimum in the enthalpy at a
predefined pressure, and the procedure repeated until the lowest
enthalpy structures have been found several times.  Such random
searching is largely unbiased, but it can often be made much more
efficient by applying constraints.  Any reasonable structure of
TeO$_2$ will contain only Te-O bonds, and therefore we have performed
most of the searches by placing O-Te-O molecules within the cells,
rather than separate atoms.  Another constraint we have employed is to
reject initial configurations in which atoms are closer than a defined
minimum separation.  We have also generated initial configurations
with the space groups which contain a specified number of symmetry
operations, and then relaxed the structures while maintaining the
symmetry.
 
Our DFT calculations were performed using the CASTEP\cite{castep}
plane wave code with the Perdew-Burke-Ernzerhof (PBE) generalised
gradient approximation (GGA) exchange-correlation
functional\cite{PerdewBE96} and ultrasoft
pseudopotentials.\cite{Vanderbilt90} For the searches we used a plane
wave cut off energy of $490$ eV and a Monkhorst-Pack\cite{monkpack}
Brillouin Zone sampling grid spacing of $2\pi \times 0.07$ \AA$^{-1}$.
All of the results reported in this paper were obtained by refining
the structures obtained in the searches and calculating their
properties using a higher level of accuracy consisting of a plane wave
cut off energy of 800 eV and a grid spacing of $2\pi \times 0.03$
\AA$^{-1}$.
 
We first performed searches at 150 GPa.  Unconstrained searches were
performed using $2$ and $4$ formula units of TeO$_2$.  Another set of
searches was performed using initial configurations built by applying
the symmetry operations of space groups chosen randomly from those
with $n$ operations to the randomly chosen positions of a Te atom and
two O atoms, with $n = $ $3$, $4$, $6$ and $8$, all subject to a
minimum separation of $r_{\rm min} = 1.3$ {\AA}. Searches were then
performed using O-Te-O molecules with initial bond angles of
$120^{\circ}$, with $1$, $2$, and $3$ molecular units and space groups
with $n=4$ operations, again with $r_{\rm min} = 1.3$ {\AA}.
Additional searches were performed at 280 GPa using molecules with
initial bond angles of $120^{\circ}$.  We used $1$ molecular unit with
$n=4$ symmetry operations, $3$ molecular units and $n=2$ symmetry
operations and $r_{\rm min} = 1.3$ {\AA}, and $4$ molecular units with
$n=2$ symmetry operations and $r_{\rm min} = 1.2$ {\AA}.  The searches
produced a total of about 1800 relaxed structures.

\begin{figure}  
\includegraphics*[width=0.5\textwidth]{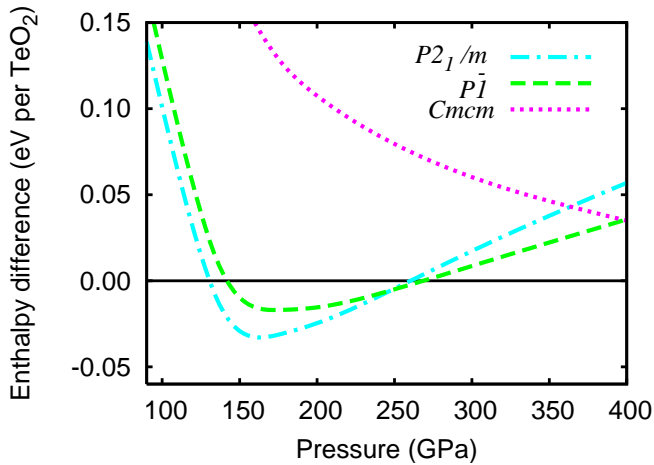}  
\caption{\label{fig:h_vs_p} (Color online) Enthalpy per TeO$_2$ unit 
relative to that of the cotunnite structure, as a function of 
pressure.} 
\end{figure}

\section{Results from structure searching}  
 
\begin{figure}  
\includegraphics*[width=0.5\textwidth]{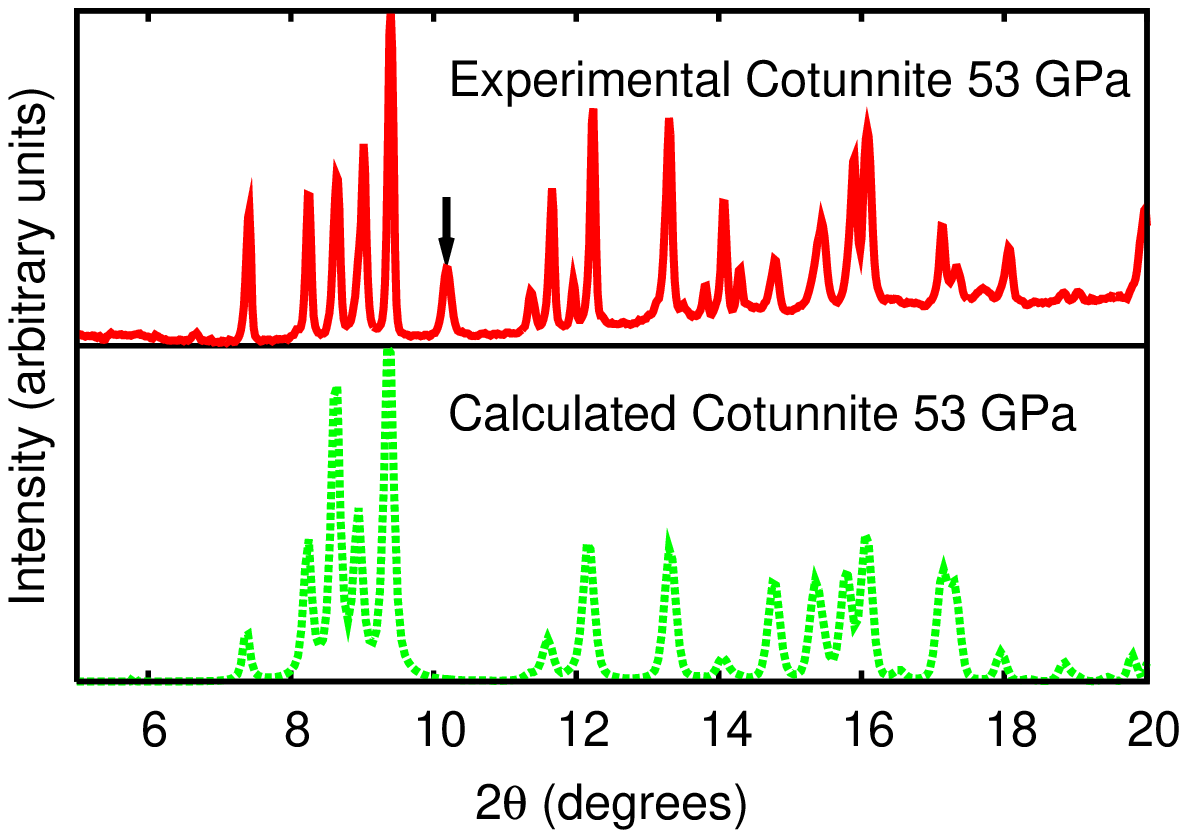} 
\includegraphics*[width=0.5\textwidth]{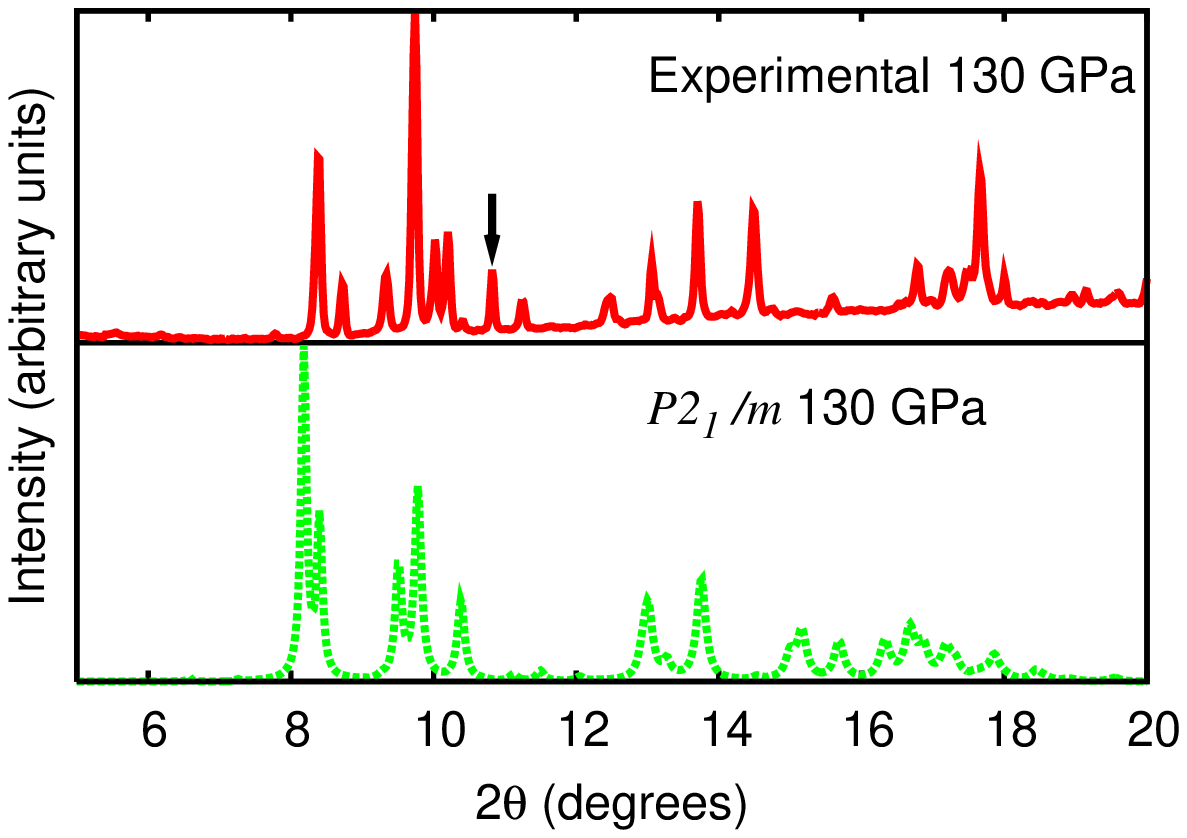}  
\caption{\label{fig:xrays} (Color online) Comparison of observed (red
solid lines) and calculated (green dashed lines) x-ray diffraction
data for the cotunnite structure (upper box) at $53$ GPa, and the
$P2_1/m$ structure (lower box) at $130$ GPa.  The experimental
diffraction data is from Ref.\ \cite{Sato}.  The experimental and
calculated data were obtained with an x-ray wavelength of
$\lambda=0.4254$ {\AA}.  The black arrows mark known impurity lines in
the experimental data,\cite{Sato} although other impurity lines may
also be present.}
\end{figure}

\begin{figure}  
\includegraphics*[width=0.4\textwidth]{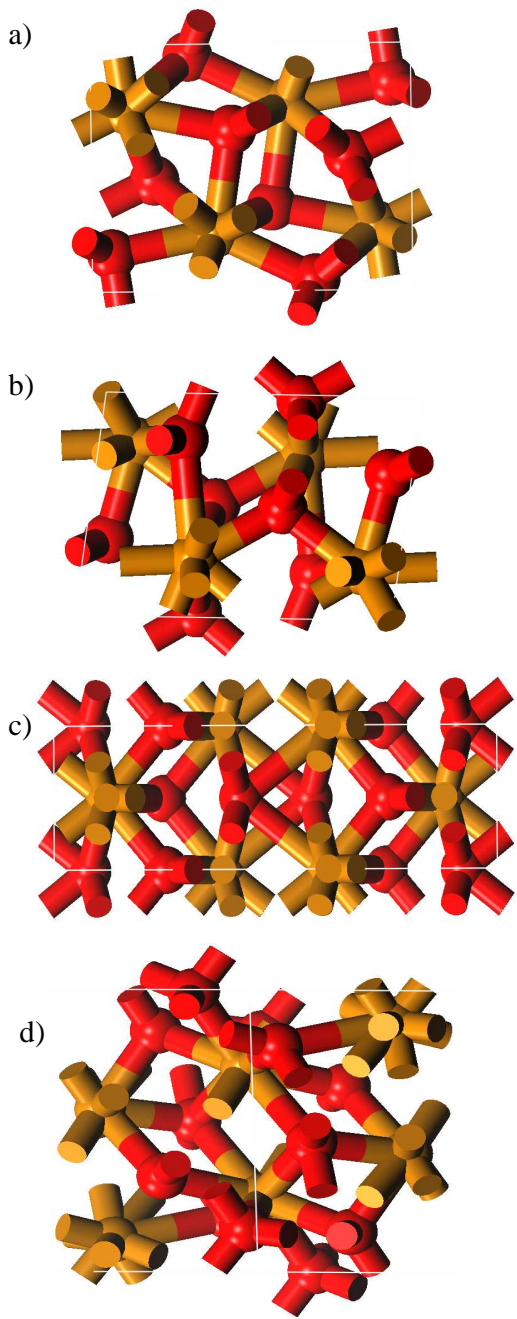}  
\caption{\label{fig:structures} (Color online) $a)$ Cotunnite, $b)$ 
$P2_1/m$, $c)$ $Cmcm$ and $d)$ $P\bar{1}$ structures. The yellow 
(light) spheres are Te atoms and the red (dark) spheres are O atoms.} 
\end{figure}

Enthalpy-pressure curves for the more stable phases are shown in Fig.\
\ref{fig:h_vs_p}.  A structure of $P2_1/m$ symmetry was consistently
the lowest enthalpy phase found at $150$ GPa in all searches with a
total of $4$ formula units, and also in the $8$ unit search with $2$
symmetry operations applied to $4$ molecular units. The structure with
the second lowest enthalpy in these searches was always found to be
the $Pnma$ cotunnite structure.  Searches with $2$ and $3$ formula
units did not yield structures with enthalpies as low as the searches
with $4$ or more, but a search with $6$ formula units produced a
low-symmetry $P\bar{1}$ structure which has an enthalpy between that
of cotunnite and $P2_1/m$ at $150$ GPa.  The cotunnite, $P2_1/m$,
$P\bar{1}$ and $Cmcm$ structures are shown in Fig.\
\ref{fig:structures} and their structural parameters are reported in
Table \ref{table:structures}. The $12$ unit search did not reveal any
new structures that were lower in enthalpy than those already
mentioned, although a previously unseen and fairly-low-enthalpy
structure with space group $P2_1/c$ was found.  The searches at 280 GPa
did not yield any further low-enthalpy structures.

\begin{table*}  
\begin{ruledtabular}  
\begin{tabular}{clllllll}  
Space group      & \multicolumn{3}{c}{Lattice parameters}          & \multicolumn{4}{c}{Atomic coordinates} \\  
                 & \multicolumn{3}{c}{(\AA, $^{\circ}$)}            & \multicolumn{4}{c}{(fractional)}    \\\hline  
$Pnma$           & $a$=4.927      & $b$=3.223     & $c$=6.389      & Te1 & 0.2398 & 0.2500 & 0.6104       \\  
                 & $\alpha$=90.00 & $\beta$=90.00 & $\gamma$=90.00 & O1  & 0.1536 & 0.2500 & 0.9378       \\  
                 &                &               &                & O2  & 0.0482 & 0.2500 & 0.3035       \\  
$P2_1/m$         & $a$=6.287      & $b$=3.577     & $c$=4.475      & Te1 & 0.1131 & 0.7500 & 0.2227       \\  
                 & $\alpha$=90.00 & $\beta$=97.15 & $\gamma$=90.00 & Te2 & 0.3503 & 0.2500 & 0.7361       \\ 
                 &                &               &                & O1  & 0.0711 & 0.7500 & 0.6543       \\  
                 &                &               &                & O2  & 0.2638 & 0.2500 & 0.2032       \\  
                 &                &               &                & O3  & 0.3930 & 0.7500 & 0.4901       \\  
                 &                &               &                & O4  & 0.3821 & 0.7500 & 0.9774       \\  
$Cmcm$           & $a$=3.014      & $b$=10.144    & $c$=3.232      & Te1 & 1.0000 & 0.8806 & 0.7500       \\  
                 & $\alpha$=90.00 & $\beta$=90.00 & $\gamma$=90.00 & O1  & 1.0000 & 0.7527 & 0.2500       \\  
                 &                &               &                & O2  & 1.0000 & 0.5783 & 0.7500       \\  
$P\bar 1$        & $a$=4.473      & $b$=5.963     & $c$=6.280      & Te1 & 0.1777 & 0.8322 & 0.1146       \\  
                 & $\alpha$=99.95 & $\beta$=97.56 & $\gamma$=111.84& Te2 & 0.5173 & 0.4923 & 0.2671       \\ 
                 &                &               &                & Te3 & 0.8648 & 0.1658 & 0.4231       \\  
                 &                &               &                & O1  & 0.1107 & 0.1638 & 0.1524       \\  
                 &                &               &                & O2  & 0.0904 & 0.5140 & 0.3393       \\  
                 &                &               &                & O3  & 0.4251 & 0.1584 & 0.4611       \\  
                 &                &               &                & O4  & 0.6011 & 0.1671 & 0.1441       \\  
                 &                &               &                & O5  & 0.7083 & 0.8322 & 0.1993       \\  
                 &                &               &                & O6  & 0.7564 & 0.5003 & 0.0060       \\        
\end{tabular}  
\end{ruledtabular}  
\caption{\label{table:structures} {Structures of the cotunnite 
($Pnma$, $Z=4$ formula units per primitive cell), $P2_1/m$ ($Z=4$ 
formula units per primitive cell), $P\bar 1$ ($Z=6$ formula units per 
primitive cell), and $Cmcm$ ($Z=6$) phases of TeO$_2$ at 130 GPa.}} 
\end{table*}  
 
In our experience, the appearance of a very low symmetry structure,
such as $P\bar{1}$, as a low enthalpy phase suggests that another
structure of even lower enthalpy might exist.  We therefore performed
an additional type of search using the cell obtained by doubling that
of $P\bar{1}$ in each direction, giving a cell containing 48 formula
units.  We then performed ``shakes'' of the larger structure in which
all atoms were displaced in random directions by a distance chosen
randomly between 0 and 0.25 \AA, and then relaxed, but in each case the original
$P\bar{1}$ structure was recovered.
 
The transition from cotunnite to $P2_1/m$ occurs at $130$ GPa in our
calculations.  Sato \emph{et al.}\ observed a phase transition at $80$
GPa after heating the sample to $1000$ K, and at $100$ GPa at room
temperature.  Heating helps in overcoming kinetic barriers which are
expected to be large in oxides and can also help to reduce anisotropic
stresses.  A temperature of $1000$ K could, however, affect the
coexistence pressure.  The agreement between the measured transition
pressure and the theoretical coexistence pressure is satisfactory,
given the uncertainty in the experimental transition pressure and the
fact that our calculations are at zero temperature.  The maximum
stabilization of the $P2_1/m$ phase over cotunnite is about 0.031 eV
per TeO$_2$ unit at 175 GPa, which is quite small but easily resolved
in our calculations.  Such small enthalpy differences are often given
quite accurately in DFT calculations for $sp$ bonded materials where
the volumes and the nature of the inter-atomic bonding in the two
phases are very similar, as is the case here.
 
\begin{figure}  
\includegraphics*[width=0.5\textwidth]{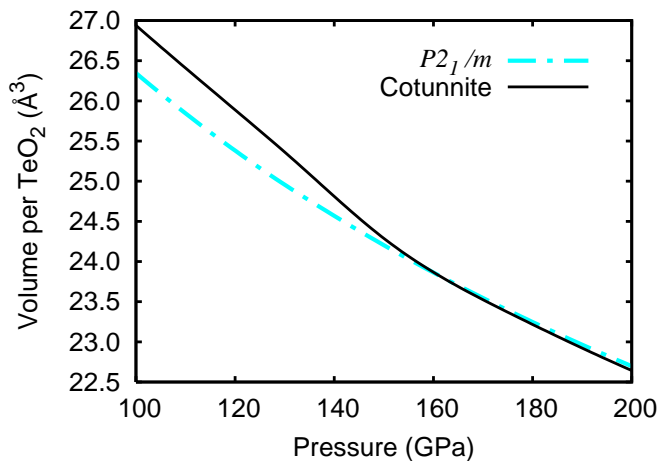}  
\caption{\label{fig:vol_plot} (Color online) Volume per TeO$_2$ 
formula unit of the cotunnite and $P2_1/m$ structures.} 
\end{figure}

Fig.\ \ref{fig:h_vs_p} shows that the $P\bar{1}$ structure is
marginally the most stable in the pressure range 248--269 GPa,
although the enthalpies of the $P\bar{1}$, $P2_1/m$, and cotunnite
phases differ by less than 0.0023 eV per formula unit in this range.
The cotunnite structure becomes more stable than the $P\bar{1}$ and
$P2_1/m$ structures again at around $260$ GPa.  This re-entrant
behavior of the cotunnite structure is quite unexpected.  The origin
of the apparent `kink' in the enthalpies of the other structures
relative to cotunnite in Fig.\ \ref{fig:h_vs_p} actually lies with the
nature of the cotunnite structure itself, as highlighted in Fig.\
\ref{fig:vol_plot}.  At pressures up to about $160$ GPa, the
compressibility of the cotunnite phase is nearly constant and is
larger than that of $P2_1/m$, but at higher pressures the
compressibilities are similar.  The region of high compressibility of
the cotunnite structure is predominantly associated with compression
along the $a$ axis, whilst the $c$ axis actually increases in length
from $125$ to $150$ GPa, before continuing to decrease steadily with
increasing pressure.  The cotunnite phase has a larger volume
at pressures below about $160$ GPa, but it is slightly smaller at
higher pressures, which tends to favor cotunnite over $P2_1/m$.

The theoretical and experimental diffraction data for the cotunnite
structure shown in Fig.\ \ref{fig:xrays} are in very good agreement.
The discrepancies in relative peak heights might arise from the form
factors used to generate the theoretical data, from the differences in
structures due to the approximate DFT and from the lack of temperature
effects in the theoretical structure.  The black arrow indicates an
impurity line in the experimental data identified by Sato \emph{et
al.},\cite{Sato} which we see is absent in the theoretical data.  The
level of agreement between the theoretical and experimental x-ray data
for the cotunnite phase gives us a benchmark for making a similar
comparison for the post-cotunnite phase.  The good level of agreement
between the theoretical and experimental data for the post-cotunnite
phase shown in Fig.\ \ref{fig:xrays} lends strong support to the
viability of the $P2_1/m$ structure as a candidate for the
post-cotunnite phase of TeO$_2$.  We note that the impurity line
indicated by a black arrow in Fig.\ \ref{fig:xrays} for the
post-cotunnite phase is absent in the theoretical data.  Sato \emph{et
al.}\cite{Sato} comment that other peaks in the experimental data may
also be impurity lines, which could explain why some of the peaks 
are missing in the theoretical data.
 
The CN of the $Cmcm$ and $P2_1/m$ structures are both found to be ten,
in comparison with cotunnite which has CN=9. The average of the nine
nearest-neighbour Te-O distances in the cotunnite structure at 130 GPa
is 2.16 {\AA} with a range of 2.02--2.28 {\AA}. For $P2_1/m$ the
average is 2.23 {\AA} with a significantly larger distribution of bond
lengths of 1.98--2.99 {\AA}.  The $Cmcm$ structure has Te-O bond
lengths of 2.02--2.51 {\AA} with an average of 2.22 {\AA}, slightly
less than for $P2_1/m$ due to the smaller distortions of the oxygen
polyhedra in $Cmcm$.  The oxygen coordination numbers with respect to
neighbouring Te atoms are $4$ and $5$ in cotunnite and $4$ and $6$ in
$Cmcm$ and $P2_1/m$.

The $P2_1/m$ structure was studied in several other dioxides to
establish whether it might be a more general post-cotunnite
phase. TiO$_2$, PoO$_2$, ThO$_2$, SeO$_2$, SiO$_2$, and HfO$_2$ were
tested, but no evidence was found to suggest that $P2_1/m$ is more
stable than cotunnite in any of these materials.

\section{Electronic structure of the phases}  
  
\begin{figure}  
\includegraphics*[width=0.5\textwidth]{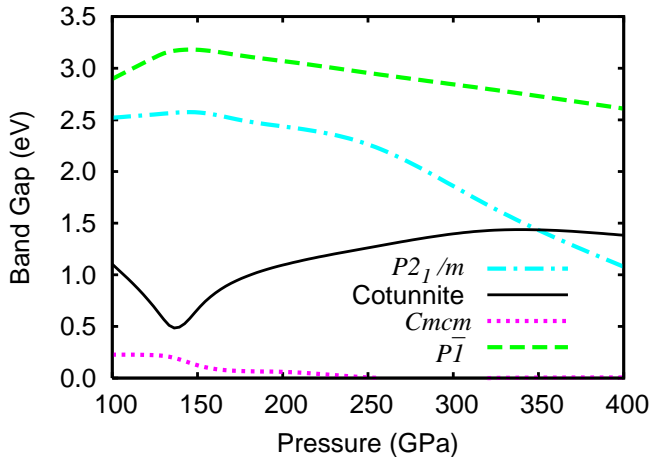}  
\caption{\label{fig:bg_plot} (Color online) Band gaps of the
cotunnite, $P2_1/m$, $Cmcm$ and $P\bar{1}$ structures as a function of
pressure. }
\end{figure}

The pressure dependence of the calculated band gaps of the structures
are shown in Fig.\ \ref{fig:bg_plot}. Above about $135$ GPa, the band
gaps of the $P2_1/m$, $Cmcm$, and $P\bar{1}$ structures decrease with
increasing pressure, however the cotunnite band gap unexpectedly
begins to increase sharply from a minimum of $0.49$ eV before
levelling off at higher pressures.  This kink in the pressure
dependence of the band gap of cotunnite approximately coincides with
the change in compressibility seen in Fig.\ \ref{fig:vol_plot}.  The
band gap of the $Cmcm$ structure falls almost to zero by $250$ GPa,
although increasing the pressure further does not lead to overlapping
valence and conduction bands. The insulating nature of the $P2_1/m$
phase is in agreement with the experimental observation of Sato
\emph{et al.}\cite{Sato} that the material was not opaque up to the
highest experimental pressure of $150$ GPa.
 
\begin{figure}  
\includegraphics*[width=0.5\textwidth]{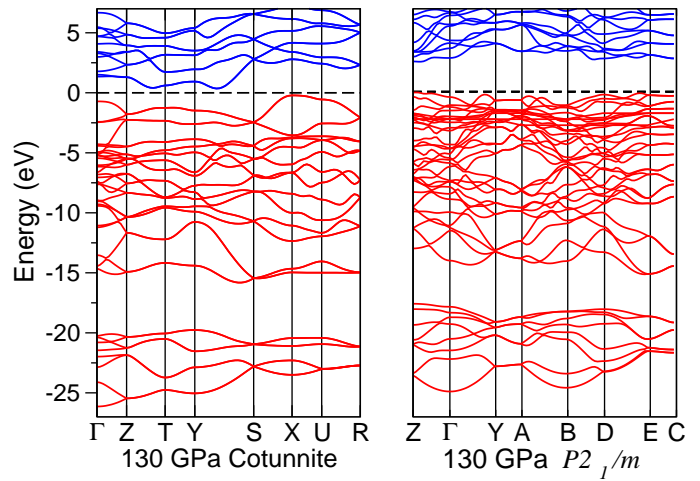}  
\caption{\label{fig:bs} (Color online) Band structures of the
cotunnite (left) and $P2_1/m$ phases (right) at $130$ GPa. The bands
with energies below zero are occupied and those at higher energies are
unoccupied.}
\end{figure}

The calculated band structures of the cotunnite and $P2_1/m$ phases at
$130$ GPa are shown in Fig.\ \ref{fig:bs}. The much larger band gap of
the $P2_1/m$ phase is apparent.  The eight lowest-energy bands for
both phases arise from the O$2s$ states, and these lie at higher
energies for the $P2_1/m$ structure, which has a somewhat smaller
occupied bandwidth than cotunnite.  Note that the band structure
calculations were performed at the PBE-GGA level and are therefore
expected to underestimate the true band gaps.

\section{Conclusions} 
 
We have searched for the post-cotunnite phase of TeO$_2$ using the
AIRSS method. Our study supports the experimental observation of a
post-cotunnite phase of TeO$_2$ at pressures readily accessible within
a diamond anvil cell.  We predict a transition to a tenfold
coordinated $P2_1/m$ phase at 130 GPa (at zero temperature), for which
the calculated x-ray diffraction data are in good agreement with
experiment.  Although the $P2_1/m$ phase has a smaller volume than the
cotunnite phase up to about 160 GPa, cotunnite has a slightly smaller
volume at higher pressures, and we predict that the cotunnite phase
re-enters at about 260 GPa.  The $P2_1/m$ phase does not appear to be
a general post-cotunnite phase for the dioxides.  The $P2_1/m$ phase
is found to be an insulator over the range of pressures studied, up to
$400$ GPa, and hence should not appear opaque, in agreement with
experiment.\cite{Sato} Higher quality x-ray diffraction data are
required to confirm whether our assignment of the $P2_1/m$ structure
to the post-cotunnite phase of TeO$_2$ is correct.

\section{Acknowledgements} 
 
This work was supported by the Engineering and Physical Research
council (EPSRC) of the UK. Computational resources were provided by
the Cambridge High Performance Computing Service.  We thank the
authors of Ref.\ \cite{Sato} for providing their x-ray diffraction
data in numerical form.

\end{document}